\documentclass[prd,preprintnumbers,nofootinbib,aps]{revtex4}
\usepackage{graphicx}
\usepackage{dcolumn}
\usepackage{bm}
\usepackage{epsfig,amsmath}
\usepackage{amssymb}
\usepackage{subfigure}
\usepackage{float}
\usepackage{ulem}
\usepackage[usenames,dvipsnames]{color}
\usepackage[pagebackref=false, colorlinks=true]{hyperref}
\definecolor{redish}{rgb}{0.7,0.2,0.0}  
\definecolor{bluish}{rgb}{0.2,0.5,0.8}

\hypersetup{linkcolor=redish,          
                  citecolor=blue,        
                  filecolor=magenta,      
                  urlcolor=bluish}          

\DeclareFontFamily{U}{rsfs}{}         
\DeclareFontShape{U}{rsfs}{m}{n}{<5> rsfs5 <6><7> rsfs7          %
  <8><9><10><10.95><12><14.4><17.28><20.74><24.88> rsfs10}{}     %
\DeclareMathAlphabet{\mathfs}{U}{rsfs}{m}{n}

\def \T{\Tmega}

\def \f{\frac}

\def \o{\omega}
\def \n{\nabla}
\def \a{\alpha}
\def \b{\beta}

\def \O{\Omega}
\def \g{\gamma}
\def \s{\sigma}
\def \e{\epsilon}

\def \d{\delta}
\def \D{\Delta}

\def \L{\Lambda}
\def \l{\lambda}
\def \th{\theta}
\def \r{\rho}

\def \g{\gamma}

\def \S{\Sigma}
\def \T{\Theta}

\newcommand{\be}{\begin{eqnarray}}
\newcommand{\ee}{\end{eqnarray}}

\begin{document}

\title{Gravitational analog of Faraday rotation in the magnetized Kerr and Reissner-Nordstr\"om spacetimes}

\author{Chandrachur Chakraborty}
\email{chandrachur.c@manipal.edu}
\affiliation{Manipal Centre for Natural Sciences,
Manipal Academy of Higher Education, Manipal 576104, India}
\affiliation{Department of Physics, Indian Institute of Science, Bengaluru 560012, India}

\begin{abstract}
It is known that the gravitational analog of the Faraday rotation arises in the rotating spacetime due to the nonzero gravitomagnetic field. In this paper, we show that it also arises in the ``nonrotating'' Reissner-Nordstr\"om spacetime, if it is immersed in a uniform magnetic field. The non-zero angular momentum (due to the presence of electric charge and magnetic field) of the electromagnetic field acts as the twist potential to raise the gravitational Faraday rotation as well as the gravitational Stern-Gerlach effect in the said spacetimes. The twisting can still exist even if the mass of the spacetime vanishes. In other words, the massless charged particle(s) immersed in a uniform magnetic field are able to twist the spacetime in principle, and responsible for the rotation of the plane of polarization of light. This, in fact, could have applications in the basic physics and the analog models of gravity. Here, we also study the effect of magnetic fields in the Kerr and Reissner-Nordstr\"om spacetimes, and we derive the exact expressions for the gravitational Faraday rotation and the gravitational Stern-Gerlach effect in the magnetized Kerr and Reissner-Nordstr\"om spacetimes. Calculating the lowest order of the gravitational Faraday effect arisen due to the presence of a magnetic field, we show that the logarithm correction of the distance of the source and observer in the gravitational Faraday rotation and gravitational Stern-Gerlach effect for the said spacetimes is an important consequence of the presence of the magnetic field. From the astrophysical point of view, our result could be helpful to study the effects of (gravito)magnetic fields on the propagation of polarized photons in the strong gravity regime of the collapsed object.
\end{abstract}

\maketitle

\section{\label{intro}Introduction}
Faraday effect is a magneto-optical phenomenon discovered by Michael Faraday in 1845. It arises due to the interaction between the light and magnetic field of the medium. If a beam of
plane polarized light is passed through a magnetic field, the plane of polarization is rotated by an angle proportional to the field intensity, which is known as the Faraday effect or Faraday rotation. Einstein's General Relativity predicts that the light rays passing a massive object bend towards it, i.e., the light is affected by the gravitational field. Not only that, it has also been shown \cite{fl, ish} that the plane of polarization of the light rays is rotated by some finite angle depending on the angular momentum of the black hole, even if there is no magnetic field. This means that one can see the gravitational analog of the Faraday rotation in a rotating spacetime, and the gravitomagnetic field (gravitational analog of the magnetic field) is responsible for this phenomenon. The plane of polarization of the light rays are rotated due to this gravitomagnetc field. Therefore, this {\it new} Faraday rotation is called as the gravitational analog of the Faraday rotation or the so-called gravitational Faraday rotation \cite{fl, ish, nz, fr1}. Note that, the gravitational Faraday rotation is not related to the magnetic field. It is only related to the gravitomagnetic field, i.e, the rotation of the spacetime. One cannot see the gravitational Faraday rotation in a non-rotating spacetime like the Schwarzschild spacetime. Thus, as of now, one can expect the graitational Faraday rotation in the Kerr spacetime. The Taub-NUT spacetime has no intrinsic rotation \cite{cm} but it has a sense of rotation due to the presence of NUT charge. However, it has been shown \cite{nz} that no gravitational Faraday rotation is occurred in the Taub-NUT spacetime. 

In a stark contrast, we show in this paper that the gravitational Faraday rotation can occur in the non-rotating 
Reissner-Nordstr\"om (RN) spacetime, if it is immersed in a uniform magnetic field. The RN spacetime is the electrovacuum solution of the Einstein-Maxwell equation with mass and electric charge only. If it is immersed in the magnetic field, its electric field and magnetic field together constitutes a non-zero electromagnetic angular momentum, i.e., it gives a {\it rotational sense} to the magnetized RN spacetime. This electromagnetic angular momentum acts as the gravitomagnetic field, and, therefore, it shows the gravitational Faraday rotation, i.e., if the light passes through this spacetime, the plane of polarization of the light or the electromagnetic wave rotates. Note that in a very recent paper, the exact solution of the magnetized Reissner-Nordstr\"om (mRN) solution was investigated by \cite{gib}.

The magnetic field plays an important role in the many astrophysical phenomena, e.g., the MHD simulation for the accretion mechanism \cite{gam}, imaging of the black hole shadow \cite{eht7, eht8, gcyl}, magnetic Penrose process \cite{dad, tur} etc. Although the magnetic Penrose process was proposed assuming the magnetic field to be asymptotically uniform \cite{dad}, the Blandford-Znajek (BZ) mechanism \cite{bz} is generally considered for the MHD simulation and to deduce the polarization of the photon ring \cite{eht7}. However,
the exact electrovacuum solution of the Einstein-Maxwell equation for the Kerr metric placed in a uniform magnetic field was first found by Wald \cite{wald}. In the next year, Ernst \cite{er} presented a general procedure for transforming an asymptotically flat axially symmetric electrovacuum solution  to an exact magnetized solution of the same. Later, the effect of a plasma in the force-free approximation was considered by Blandford and Znajek \cite{bz}.
For the lacking of the direct measurements of the exact shapes of the magnetic field configurations around a realistic collapsed objects, many other numerical techniques are used to show the strong connections between the shape of magnetosphere and the characteristics of accretion mechanism \cite{pun, mei}. In this paper, we consider the Wald \cite{wald} and/or Ernst \cite{er} solution as this is the exact electrovacuum solution of the Einstein-Maxwell equation. Secondly, the uniform magnetic field configuration assumed in the magnetic Penrose process \cite{dad, tur} seems more efficient  than the magnetic field configuration of the BZ mechanism for the electromagnetic extraction of the rotational energy from a rotating black hole.

There was a draw back of the Ernst solution: it produces the conical singularities at the polar axis \cite{ag1}, which was removed by the Ernst-Wild (see \cite{wild}) solution in order to obtain a physically meaningful solution \cite{ag2}. Later, Aliev and Galt'sov \cite{ag3} applied this solution to observe the magnetic precession (see also \cite{ccb, rp}) in black hole systems with magnetized accretion disks. It is known that the gravitational energy is much greater than the electromagnetic energy, but those are comparable if the strength of the magnetic field $(B)$ surrounding a collapsed object with mass $M$ is the order of \cite{ag2, gp}
\begin{eqnarray}
 B \simeq B_{\rm max} \sim 2.4 \times 10^{19} \f{M_{\odot}}{M}
\end{eqnarray}
where $M_{\odot}$ is the solar mass. 
The strength of the magnetic field surrounding black holes is considered much smaller than the value of $B_{\rm max}$ (i.e., $B << B_{\rm max}$) but the investigations suggest that the surrounding spacetimes around a black hole could be highly distorted for $B \sim B_{\rm max}$. Thus, this magnetic field is very important as a background field testing of the geometry around a collapsed object \cite{sha}. 

In this paper, we have chosen the magnetized Kerr and magnetized RN spacetimes to study the effect of magnetic field on the gravitational Faraday effect. In general, the Kerr black hole is considered as the most relevant from the astrophysical point of view, and it is supposed to immerse in a non-zero magnetic field. On the other hand, although the ordinary RN spacetime is spherically symmetric, the magnetized RN is very special in this sense that it becomes axisymmetric due to the presence of magnetic field (as discussed in the second paragraph of this section). These are basically the main reasons why we are interested to study the gravitational Faraday effect in these spacetimes. We also carry out the similar study for the massless charged-RN solution, which is devoted only for the theoretical purpose.
The scheme of the paper is as follows. In Sec. \ref{sec2}, we revisit the formalism of the gravitational analog of the Faraday rotation, and derive the relation of it with the gravitational anlogue of the Stern-Gerlach effect and the so-called spin precession of a test gyroscope. In Sec. \ref{sec3}, we study the effect of magnetic field on the gravitational Faraday rotation and the gravitational Stern-Gerlach effect in the magnetized Kerr spacetime. Sec. \ref{sec4} and Sec. \ref{mls} are devoted to study the effect of magnetic field on the gravitational Faraday and Stern-Gerlach effects in the mRN spacetime and the massless mRN-like spacetime, respectively. Finally, we conclude in Sec. \ref{con}.

\section{\label{sec2}Gravitational Faraday rotation in the stationary spacetime}

The general metric of a stationary spacetime can be written as 
\begin{eqnarray}
 ds^2=g_{\mu\nu}dx^{\mu}dx^{\nu} &=& g_{00}(dx^0)^2+2g_{0i} dx^0 dx^i+g_{ij}dx^idx^j
\\
 &=& h(dx^0-g_i dx^i)^2-\g_{ij}dx^idx^j
 \label{gen}
\end{eqnarray}
where 
\begin{eqnarray}
h=g_{00}, \,\,\,\,\, g_i=-\f{g_{0i}}{h}, \,\,\,\,\, \g_{ij}=-g_{ij}+\f{g_{0i}g_{0j}}{g_{00}}.
\label{gmij}
\end{eqnarray}
The Greek indices represent the time and space components, i.e., $x^{\mu}=(x^0 \equiv t, x^i)$ and the Latin indices represent only the space components, i.e., $i=1,2,3$. 
We denote 
\begin{eqnarray}
 g \equiv {\rm det}(g_{\mu\nu}), \,\,\,\,\, \g \equiv \rm{det}(\g_{ij})
\end{eqnarray}
and, hence, $-g=h \g$. 
In a static spacetime ($g_i=0$), the metric $\g_{ij}$ reduces to the so-called optical metric \cite{sa, fr1}. Spatial trajectories of light rays
are geodesics of the optical metric. Landau and Lifshitz \cite{ll} also showed that $\g_{ij}$ can be regarded as a metric of space, as opposed to spacetime. This is similar to the $(3+1)$ decomposition of the metric (Eq. \ref{gen}).
They showed that the test bodies following geodesics of spacetime depart from the geodesic of space as if acted on by the gravitational force ${\bf F}$ which can be expressed as \cite{lnbl} \footnote{We use the geometrized unit ($G=c=1$) in this whole paper except Eq. (\ref{ofe}).}.
\begin{eqnarray}
 {\bf F}=\f{m_0}{\sqrt{1-v^2}}\left({\bf E_g}+{\bf v}\times \sqrt{h}{\bf B_g} \right)
 \label{force}
\end{eqnarray}
where 
\begin{eqnarray}
 {\bf E_g}=-{\bf \n}~{\rm ln \sqrt{h}}=-\f{1}{2}\f{{\bf \n}h}{h}
\end{eqnarray}
and 
 \begin{eqnarray}
  {\bf B_g} = {\rm curl}~{\bf A} \equiv {\rm curl}~{\bf g}
  \label{bg}
 \end{eqnarray}
 are the gravitoelectric (${\bf E_g}$) and gravitomagnetic (${\bf B_g}$) fields respectively. It is needless to say here that $g_i$
 is equivalent to $A_i$, i.e., $g_i \equiv A_i$, the gravtational analog of the vector potential (of electromagnetism). Now, one can rewrite the Landau and Lifshitz's form of Einstein's equations as \cite{nz}
\begin{eqnarray}
 {\rm div}~{\bf B_g} &=& 0 ,
 \label{me1}
 \\
 {\rm curl}~{\bf E_g} &=& 0 ,
 \\
 {\rm div}~{\bf E_g} &=& -\left[\f{1}{2}(\sqrt{h}{B_g})^2+{E_g}^2 \right] ,
 \\
 {\rm curl}~(\sqrt{h}{\bf B_g}) &=& 2{\bf E_g} \times (\sqrt{h}{\bf B_g}) .
 \label{me4}
\end{eqnarray}
It would be interesting to notice that $\sqrt{h}{\bf B_g}$ also appears in the expression of force (Eq. \ref{force}), and, the right hand side of Eq. (\ref{me4}) could be considered as an energy current corresponding to the Poynting vector flux of gravitational field energy. Note, all the operations in Eqs. (\ref{me1}-\ref{me4}) are defined in the 3D space with the metric $\g_{ij}$. However, one can also define the above gravitomagnetic fields in the following covariant forms by using the timelike Killing vector of the spacetime as \cite{nz},
\begin{eqnarray}
 E_g^{\varsigma} &=& -\f{1}{2}\f{(\zeta^{\s}\zeta_{\s})^{;\varsigma}}{|\zeta|^2},
 \\
 B_g^{\varsigma} &=& -\f{1}{2}|\zeta| \zeta^{\s}  \e_{\s}{}^{\varsigma \iota \rho} \left[\left(\f{\zeta_{\r}}{|\zeta|^2}\right)_{;\iota}-\left(\f{\zeta_{\iota}}{|\zeta|^2}\right)_{;\r}\right],
\end{eqnarray}
where $\e_{\s}{}^{\varsigma \iota \rho}$ is the four dimensional antisymmetric tensor, $|\zeta|=\sqrt{h}$ and the semicolon denotes the covariant differentiation.

Now, using the analogy with the flat spacetime, we consider the plane of polarization of an electromagnetic wave consisting two 3-vectors: the wave vector ${\bf k}$ and the polarization vector ${\bf f}$. The 4-vectors corresponding to these 3-vectors are related as
\begin{eqnarray}
 k^{\s}k_{\s}=0, \,\,\,\,\, k^{\s}f_{\s}=0,\,\,\,\,\, f^{\s}f_{\s}=1, \,\,\,\,\, ({\rm with} \,\,\,\,\, \s=0,1,2,3)
\end{eqnarray}
and, both of them ($k^{\s}$ and $f^{\s}$) are parallelly-trnasported along the null geodesic \cite{ll}. It should be useful to note here that the covariant counterparts of these 3-vectors (${\bf k}$ and ${\bf f}$) are not the spatial components of the covariant 4-vetors $k_{\s}$ and $f_{\s}$, rather
\begin{eqnarray}
 ^{(3)}k_j=\g_{ij} ^{(3)} k^i=^{(4)}k_j+k_0 g_j
 \end{eqnarray}
 and 
\begin{eqnarray}
  ^{(3)}f_j=\g_{ij} ^{(3)}f^i=^{(4)}f_j+f_0 g_j.
\end{eqnarray}
There is a gauge freedom which enables us to put $f_0=0$ without the loss of generality \cite{nz}. Now, applying the above decomposition with the gauge condition and using the equations of parallel transport for $k^{\s}$ and $f^{\s}$, 
the evolution equations of ${\bf k}$ and ${\bf f}$ along the ray were derived as \cite{fl},\cite{nz}
\begin{eqnarray}
 ^3{\bf \n}_{\bf k}{\bf k} &=& {\bf L} \times {\bf k}+({\bf E_g}.{\bf k}){\bf k}
 \label{k}
 \\
 ^3{\bf \n}_{\bf k}{\bf f} &=& {\bf L} \times {\bf f}
 \label{f}
\end{eqnarray}
where 
\begin{eqnarray}
 {\bf L}=-\f{1}{2}k_0\left[{\bf B_g}-\f{1}{2}(\bf B_g.f){\bf f}+\f{1}{|{\bf f}|}{\bf E_g}.({\bf k \times f}){\bf f} \right].
\end{eqnarray}
If only the second term exists on the RHS of Eq. (\ref{k}), it would mean, by comparison with the 4D definition of the parallel transport, that the 3-vector ${\bf k}$ is parallely transported along the projection of the null geodesic,but the presence of the first term indicates that ${\bf k}$ is rotated by the angular velocity ${\bf L}$. The same rotation 
also appears for the polarization vector ${\bf f}$ (see Eq. \ref{f}). Thus, both of the equations together leads to this important fact that the polarization plane rotates with the angular velocity
${\bf L}$ along the projected null geodesic. However, Ref. \cite{nz} derived the angle of rotation ($\chi$) around the tangent vector ${\bf {\hat k}}$ along the path between the source and the observer as \cite{nz}:
\begin{eqnarray}
 \chi &=& \int_{\rm source}^{\rm observer} {\bf L}.\hat{\bf k} ~d\l
 \\
 &=& -\f{1}{2} \int_{\rm source}^{\rm observer} k_0 {\bf B_g}.\hat{\bf k} ~ d\l
 \label{eq17}
\end{eqnarray}
where $\l$ is the affine parameter along the ray. If one considers a small line element $\bf{dl}$ along the path of the ray, one can write (see Eq. 19 of \cite{nz})
\begin{eqnarray}
 \f{k_0^2}{h}=\left(\f{dl}{d\l}\right)^2 .
 \label{eq18}
\end{eqnarray}
Now, substituting Eq. (\ref{eq18}) in Eq. (\ref{eq17}) with $\hat{\bf k} dl=\bf{dl}$, one obtains from Eq. (\ref{eq17}) 
\begin{eqnarray}
 \chi &=& -\f{1}{2}\int_{\rm source}^{\rm observer}  \sqrt{h}~{\bf B_g}.{\bf dl}
 \label{o0}
 \\
 &=& -\f{1}{2}\int_{\rm source}^{\rm observer} {\rm curl} (\sqrt{h}~{\rm curl}~{\bf g}).{\bf dS}
 \label{o1}
\end{eqnarray}
\cite{nz} where ${\bf dS}$ represents the surface enclosed by the path of the light ray (i.e., a null geodesic) which passes close to a collapsed object like the black hole. \footnote{Note, Eq. (\ref{o0}) was directly applied to deduce the gravitational Faraday rotation in the `spherically symmetric' Taub-NUT spacetime considering a closed path around the Taub-NUT hole (see Sec. IV of \cite{nz}), whereas Eq. (\ref{o1}) was applied for the `axisymmetric' Kerr spacetime (see Sec. V of \cite{nz}). For the latter, the integration was performed over the orbital plane which was enclosed by the orbit of null geodesic. } Here, we use the Stokes theorem and Eq. (\ref{bg}) to obtain Eq. ({\ref{o1}}) from Eq. ({\ref{o0}}). Eq. (\ref{o0}) indicates that a light ray propagates along the line of sight starting from the source at infinity and ending at the observer. The light rays reaching the observer along the two different lines of sight \cite{nag} traverses two different paths and the rotation angles for
their planes of polarization is given by Eq. (\ref{o0}). Actually, the surface enclosed (as shown in Eq. \ref{o1}) by the two referred paths \cite{nag} could be physically referred to a spherical corona between the source and the observer location.

It was pointed out in \cite{fr2, fr1} that the spin-optical interaction described by the effective force is proportional to `${\rm curl~curl}~{\bf g}$' (see Eq. 71 and Eq. (81) of \cite{fr2}), if one use the definition of $\g$ metric as of Eqs. (2) and (6) of \cite{fr1}. In our  case, the effective force should be proportional to `$h~{\rm curl} (\sqrt{h}~{\rm curl}~{\bf g})$'. This effective force increases when a photon approaches the collapsed object and reaches its maximum near its radial turning point. 
Note, if one use the definition of the $\g$ metric of \cite{fr1}, one obtains the angle of rotation ($\bar \chi$) as 
\footnote{We follow the definition of $\chi$ mentioned in Eq. (\ref{o1}) (not of Eq. \ref{o4}) in the whole paper.
}
\begin{eqnarray}
  {\bar \chi} &=& -\f{1}{2}\int_{\rm source}^{\rm observer} {\rm curl} \left({\rm curl}~{\bf g} \right).{\bf dS} .
  \label{o4}
\end{eqnarray}

One can take the example of the Kerr spacetime. 
From the atsrophysical point of view, the most relevant spacetime  is the Kerr spacetime to describe the astrophysical collapsed objects. The Kerr metric in the Boyer-Lindquist coordinates $x^{\mu}\equiv (t,r,\th,\phi)$ can be written in the form of Eq. (\ref{gmij}) with
\begin{eqnarray}
 g_{00}=h=\left(1-\f{2Mr}{\S}\right)\, , \,\,\,\,\, {\bf g} \equiv g_{\phi}=-\f{2aMr\sin^2\theta}{\S-2Mr}
\end{eqnarray}
and
\begin{eqnarray}
\g_{ij}dx^idx^j=\f{\S}{\Delta}dr^2  
+\S d\theta^2+\f{\D }{h} \sin^2\theta d\phi^2
\label{k1}
\end{eqnarray}
where, $a$ is the Kerr parameter, defined as $a=\f{J}{M}$, 
the angular momentum ($J$) per unit mass $(M)$ and
\begin{equation}
 \S=r^2+a^2 \cos^2\theta,        \,\,\,\,\,      \Delta=r^2-2Mr+a^2 .
\label{k2}
\end{equation}

Considering Eqs. (127-131) in the notation of \cite{fr1} one can obtain from Eq. (\ref{o4})
\begin{eqnarray}
  {\bar \chi_{\rm Kerr}} &=& -\f{1}{2}\int_{\rm source}^{\rm observer} \f{4aM^2}{\S^3}.\f{\S \sin\th}{h^2}~dr d\th
  \label{o6}
  \\
 &=&  -2aM^2\int_{\rm source}^{\rm observer} \f{\sin\th}{(\S-2Mr)^2}~dr d\th
 \label{o5}
\end{eqnarray}
 which is the same as of Eq. (23) of \cite{nz}. In the weak-field regime $(r >> M)$, one can obtain from Eq. (\ref{o5}): ${\bar \chi_{\rm Kerr}} \sim aM^2/R^3$, where $R$ is the distance between the source and the observer.
In our case, the effective force for the gravitational Faraday effect changes as 
\begin{eqnarray}
\left|h~{\rm curl} (\sqrt{h}~{\rm curl}~{\bf g}) \right| =\left|h\f{4aM^2}{\sqrt{h}\S^2(\S-2Mr)}.\f{\sqrt{\D} \sin\th}{\sqrt{h}}\right| &=& \f{4aM^2 \sqrt{\D}\sin\th}{\S^2(\S-2Mr)} 
  \label{fe}
  \\
  & \approx & \f{4aM^2}{r^5} \sin\th
   \label{fek}
 \end{eqnarray}
where Eq. (\ref{fek}) is valid far away ($M/r << 1$) from a slowly-rotating ($a/M << 1$) Kerr black hole as showed earlier in \cite{fr2}.

\subsection{\label{sec2.1}Relation between the gravitational Faraday rotation and the so-called spin precession}

Considering Eq. (\ref{bg}), and using the relation (${\bf e}_{\hat{p}}={\bf e}_{p}/\sqrt{\g_{pp}}$ \cite{jh}) between the orthonormal basis vectors (${\bf e}_{\hat{p}}$) and the coordinate basis vectors (${\bf e}_{p}$), one obtains the gravitomagnetic field for the Kerr spacetime as
\begin{eqnarray}
 ({\rm curl}~{\bf g})^{\hat{p}}~{\bf e}_{\hat{p}}=-2aM\left[\f{2r\sqrt{\D}\cos\th}{\S(\S-2Mr)^{3/2}}~{\bf e}_{\hat{r}} 
+\f{(r^2-a^2\cos^2\th)\sin\th }{\S(\S-2Mr)^{3/2}}~{\bf e}_{\hat{\th}}\right].
\label{l2}
\end{eqnarray}
Interestingly, if one multiplies Eq. (\ref{l2}) with `$-\sqrt{h}/2$', one obtains the same expression (Eq. (42) of \cite{cm}) which was obtained as the spin precession frequency ($\O_ s$), or, so-called the Lense-Thirring (LT) precession \cite{lt} frequency of a test gyroscope or a test spin (for the detailed spin precession formalism, see \cite{ckj,ckp,koch}. So, technically, Eq. (\ref{o1}) leads to the angle of 
\begin{eqnarray}
 \chi=\int_{\rm source}^{\rm observer} ({\rm curl}~{\bf \O}_s).{\bf dS}.
 \label{xos}
\end{eqnarray}

Although the general expression for the LT precession frequency of a test spin in terms of the coordinate basis vectors was obtained in \cite{ns} as
\begin{eqnarray}
 {\bf \O} =  -\f{g_{00}}{2\sqrt {-g}} \e_{ijp} g_{i,j} \left({\bf e_{p}} + 
g_p {\bf e_{0}}\right) 
\label{s1}
\end{eqnarray}
for a general stationary spacetime, we should only consider
\begin{eqnarray}
 {\bf \O}_ s =  -\f{1}{2} \sqrt{\f{h}{\g}}~\e_{ijp}~ g_{i,j}~ {\bf e_{p}} 
 \equiv -\f{1}{2}\sqrt{h}~({\rm curl}~{\bf g})
\label{s2}
\end{eqnarray}
as we deal here only with the $\g_{ij}$ metric. $\e_{ijp}$ is the Levi-Civita symbol.

Note, in case of the Taub-NUT spacetime \cite{cbgm, cbgm2}, $({\rm curl}~{\bf g})$ comes as non-zero \cite{nz,cm}, and, hence the spin of a test gyro can precess \cite{cm} in this spacetime. On the other hand, $\left({\rm curl}~(\sqrt{h}~{\rm curl}~{\bf g}) \right)$ vanishes, and, therefore, no gravitational Faraday rotation  is induced in the Taub-NUT spacetime \cite{nz}. This example could be helpful to differentiate between the spin precession and the gravitational Faraday rotation.

\subsection{\label{faro}Angular separation $(\T)$ of the right and left circularly polarized beams due to the gravitational Faraday rotation: Gravitational Stern-Gerlach effect}

There exists a gravitational analog of the Stern-Gerlach effect \cite{fr2}, i.e., in the spacetime of a rotating collapsed object, the trajectories of the circularly polarized photons depend on their polarization. Using this analogy,
Mashhoon \cite{mas1, mas2} showed that the photons of the opposite (right and left) circular polarization emitted by a distant source deflects to the directions with the separation angle $(\T)$ 
\begin{eqnarray}
 \T \sim \f{aM}{\o D^3}
\end{eqnarray}
after scattering, where $\o$ is the photon frequency and $D$ is the distance from the photon to the body at the moment of their minimal separation. As the gravitomagnetic field depends upon position, there exists a gravitomagnetic Stern-Gerlach force $-\nabla ({\bf \O}_s . {\bf N})$ on a spinning particle with `intrinsic' spin vector ${\bf N}$. This force naturally leads to a differential deflection of the polarized beams \cite{mas3}.

In a recent paper, \cite{fr1} has studied in detail how the polarization of photons affects their motion in a
gravitational field created by a rotating massive collapsed object, and shown that the angular separation $(\T)$ of the right and left circularly polarized beams is deduced using the dimensionless parameter: $\varepsilon =\pm (M\o)^{-1}$ \cite{fr2} (with $|\varepsilon| << 1$). Thus, we obtain the relation between $\T$, $\chi$ and ${\bf \O}_s$ as
\begin{eqnarray}
 \T=\pm \f{1}{M\o }\chi &=& \mp \f{1}{2M\o}\int_{\rm source}^{\rm observer} {\rm curl} (\sqrt{h}~{\rm curl}~{\bf g}).{\bf dS} 
 \label{fra}
 \\
 &=& \pm \f{1}{M\o}\int_{\rm source}^{\rm observer} ({\rm curl}~{\bf \O}_s).{\bf dS}
\end{eqnarray}
where and $+$ and $-$
correspond to the right and left circular polarizations, respectively. Eq. (\ref{fra}) depends on $\o$, and this, in fact, can be used as the final expression to obtain the angular separation of the right and left circularly polarized beams due to the gravitational Faraday effect in a stationary spacetime. 
In the next two sections, we study the effects of the gravitational analog of Faraday rotation and Stern-Gerlach effect in the magnetized Kerr and Reissner-Nordstr\"om spacetimes. 

\section{\label{sec3}Gravitational Faraday rotation in the magnetized Kerr spacetime}

\subsection{\label{sec3.1}Brief discussion on the Kerr spacetime immersed in the uniform magnetic field}
The exact electrovacuum solution of the Einstein-Maxwell equation for the  magnetized Kerr spacetime is written as \cite{ag1, ag2}
\begin{eqnarray}
 ds^2=\left( \f{\D}{A}dt^2-\f{dr^2}{\D}-d\th^2 \right)\S|\L|^2-\f{A\sin^2 \th}{\S|\L|^2}\left(|\L_0|^2 d\phi-\varpi dt \right)^2
 \label{kerrm}
\end{eqnarray}
where 
\begin{eqnarray}
  \D=r^2+a^2-2Mr \, , \,\,\,\,\, \S=r^2+a^2\cos^2\th ,
\end{eqnarray}
\begin{eqnarray}
A=(r^2+a^2)^2-\D a^2 \sin^2\th \, , \,\,\,\,\, \varpi=\f{v -w \D}{r^2+a^2}.
 \label{ob}
\end{eqnarray}
$\L (r,\th)$ is a complex quantity and it has two parts, the real part of $\L$: ${\rm Re}~\L$ and the imaginary part of $\L$: ${\rm Im}~\L$. So, one can express it as:
\begin{eqnarray}
 \L \equiv \L (r,\th) &=& {\rm Re} \L + i~{\rm Im} \L \nonumber
 \\
 &=&1+\f{B^2\sin^2\th}{4}\left[\left(r^2+a^2\right)+\f{2a^2Mr\sin^2\th}{\S} \right]-i.\f{aB^2M\cos\th}{2}\left(3-\cos^2\th+\f{a^2\sin^4\th}{\S}\right) \nonumber
 \\
 \label{lambda}
\end{eqnarray}
where $i~(\equiv \sqrt{-1})$ represents the imaginary unit.
In the expression of $\varpi$ (Eq. \ref{ob}), 
\begin{eqnarray}
 v &=& a(1-a^2M^2B^4) 
 \\
 {\rm and,} && \nonumber
 \\
 w &=& \f{a\S}{A}+\f{aMB^4}{16}
 \left(-8r\cos^2\th(3-\cos^2 \th)-6r\sin^4\th+\f{2a^2\sin^6\th}{A}[2Ma^2+r(a^2+r^2)] \right. \nonumber
 \\
&+& \left. 
\f{4Ma^2\cos^2\th}{A}\left[(r^2+a^2) (3-\cos^2 \th)^2-4a^2\sin^2\th \right] \right)
\end{eqnarray}

It was first pointed out in \cite{his} that `magnetic' transformation of the Kerr spacetime is only locally valid, as it produces the conical singularities at the polar axis. This conical singularities on the polar axis generate some singular stress energy tensor on the right hand side of the Einstein equation in addition to the Maxwellian term. This deficiency can be removed by changing the interval of variation of the azimuthal angle $\phi$ from $2\pi$ to $2\pi |\L_0|^2$ \cite{ag1, ag2}, where
\begin{eqnarray}
|\L_0|^2=|\L(r,0)|^2=1+a^2M^2B^4
\end{eqnarray}
is the Harrison-Ernst function $\L(r,\th)$ at the polar axis $\th=0$.
That is the reason, $|\L_0|^2$ appears just before $d\phi$ in Eq. (\ref{kerrm}). Without the factor $|\L_0|^2$ before $d\phi$, the magnetized Kerr metric is not really a solution of the Einstein-Maxwell
equations elsewhere, as it produces the Ricci tensor singular at the polar axis \cite{ag1, ag2}. Such a singularity is similar to that of the cosmic string \cite{vil}, which has positive energy density and negative tension by analogy with the physical string. Note, the horizon radii 
\begin{eqnarray}
 r_{\pm}=M \pm \sqrt{M^2-a^2}
\end{eqnarray}
are same for the ordinary Kerr metric and the magnetized Kerr metric, as it do not depend on the value of $B$. On the other hand, the ergoradii in the magnetized Kerr metric depend on the value of $B$. Therefore, the expressions of ergoradii for the magnetized Kerr metric cannot be same with the ordinary Kerr metric.

\subsection{\label{sec3.2}Effective force for the  gravitational Faraday rotation in the magnetized Kerr spacetime}
For the magnetized Kerr spacetime, one can deduce 
\begin{eqnarray}
 g_{00}=h=\left(\f{\D \S |\L|^2}{A}-\f{A\varpi^2\sin^2\th}{\S |\L|^2}\right)\, , \,\,\,\,\, {\bf g} \equiv g_{\phi}=-\f{\varpi A^2 |\L_0|^2 \sin^2\theta}{\D\S^2|\L|^4-A^2\varpi^2\sin^2\th}
\end{eqnarray}
and
\begin{eqnarray}
\g_{ij}dx^idx^j=\f{\S |\L|^2}{\Delta}dr^2  
+\S |\L|^2 d\theta^2+\f{ \D |\L_0|^4}{h} \sin^2\th d\phi^2
\label{mgk}
\end{eqnarray}
from Eq. (\ref{kerrm}). Now, using Eq. (\ref{mgk}), we obtain
\footnote{Although we have calculated the exact expressions for $\left({\rm curl}~(\sqrt{h}~{\rm curl}~{\bf g}) \right)$ and $({\rm curl}~{\bf g})$, we do not show it here as those expressions are very big in size and useless for the current purpose of this paper. Those expressions could be important for the numerical calculations which we suppose to report in future. As the magnetic energy is much less than the gravitational energy for a massive magnetized collapsed object (see Sec. \ref{intro}), considering the order of magnetic field upto $B^2$ should suffice for the current purpose of this paper. Hence, we show all the required expressions upto the order of $B^2$.}
\begin{eqnarray}
 ({\rm curl}~{\bf g})^p &=& -\f{4aMr\D \cos\th}{\S^{3/2}(\S-2Mr)^{3/2}}. \nonumber
\left(1- \f{B^2 \sin^2\th}{32\S(\S-2Mr)} C_1(r,\th) \right) \d^p_r \nonumber
 \\
 &-&\f{2aM\sin\th}{\S^{3/2}(\S-2Mr)^{3/2}}\left((r^2-a^2\cos^2\th)+ \f{B^2 \sin^2\th}{128\S(\S-2Mr)}C_2(r,\th) \right) \d^p_{\th} + \mathcal{O}(B^3) \nonumber
 \\
\label{cgk}
\end{eqnarray}
and, 
\begin{eqnarray}
 \left({\rm curl}~(\sqrt{h}~{\rm curl}~{\bf g})\right)^p &=& \left[\f{4aM^2}{\S^{3/2}(\S-2Mr)^{3/2}}-\f{3aMB^2}{64\S^{5/2}(\S-2Mr)^{5/2}} C_3(r,\th)+\mathcal{O}(B^3)\right] \d^p_{\phi} \nonumber
 \\
 \label{cck}
\end{eqnarray}
where $C_1(r,\th)$, $C_2(r,\th)$ and $C_3(r,\th)$ are the functions of $r$ and $\th$. \footnote{As the exact expressions of $C_{1-5}(r,\th)$ are not very relevant for this paper, we do not mention it here. However, the exact expressions of the same can be available upon request.}

Now, one can calculate the effective force for the gravitational Faraday rotation that varies as 
\begin{eqnarray}
 \left|h~{\rm curl} (\sqrt{h}~{\rm curl}~{\bf g}) \right| =\f{4aM^2 \sqrt{\D}\sin\th}{\S^2(\S-2Mr)}-\f{aMB^2\sqrt{\D}\sin\th\ C_4(r,\th)}{64\S^3(\S-2Mr)^2}+\mathcal{O}(B^3)
  \label{femk}
 \end{eqnarray}
 where $C_4(r,\th)$ is the function of $r$ and $\th$.

The second term in the right hand side of Eq. (\ref{femk}) appears as the correction term due to the presence of magnetic field. Far away $(M/r << 1)$ from a slowly rotating Kerr black hole $(a/M << 1)$, Eq. (\ref{femk}) falls down as
\begin{eqnarray}
 \left|h~{\rm curl} (\sqrt{h}~{\rm curl}~{\bf g}) \right|_{\left(a/M << 1, M/r << 1 \right)} \approx \sin\th \left[\f{4aM^2}{r^5}-\f{3aMB^2}{r^2}(1+3\cos2\th) \right].
 \label{fes}
\end{eqnarray}
Eq. (\ref{fes}) shows that although the effective force for the gravitational Faraday effect decreases as $\sim r^{-5}$ for the ordinary Kerr spacetime (see Eq. 81 of \cite{fr2}), the same due to the presence of non-zero magnetic field decreases as $\sim r^{-2}$ in the magnetized Kerr spacetime. Interestingly, one cannot see the effect of magnetic field in Eq. (\ref{fes}) for $\th=\f{\pi}{2}-\f{1}{2}\cos^{-1}\f{1}{3} \approx 54.74^{\circ}$, as the second term in the square bracket of Eq. (\ref{fes}) vanishes for that value.

\subsection{\label{sec3.3}Logarithm correction in the gravitational Faraday rotation and Stern-Gerlach effect for the slowly-rotating magnetized Kerr black hole}

The gravitational Faraday effect in the Kerr spacetime was studied earlier \cite{ple, nz}. 
It was shown that when a light ray passes through the outside of a rotating matter, its polarization plane rotates \cite{ish}. In this section, we consider a general orbit (following \cite{nz}) around a magnetized Kerr black hole, which intersects the equatorial plane ($\th=\pi/2$) and is symmetric about it (see Appendix \ref{app2}). Using Eq. (\ref{xos}) we obtain that the polarization plane is rotated by an angle
\begin{eqnarray}
 \chi=-\int_{r_{\rm orb}(\th)}^{r_o} \int_{\th_s}^{\th_o} \left[\f{2aM^2}{(\S-2Mr)^2}-\f{aMB^2}{32(\S-2Mr)^3}C_5(r,\th) \right] \sin\th dr d\th 
 \label{ki}
\end{eqnarray}
where, $C_5(r,\th)$ is the function of $r$ and $\th$. In Eq. (\ref{ki}), $r_o$ is the location of the distant observer and $r_{\rm orb}(\th)$ is the equation of the projection  (see Eq. \ref{b7}) of the orbit in the $(r,\th)$ plane. $\th_s$ and $\th_o$ are the position angles of the source and the observer, respectively. Note, Eq. (\ref{ki}) reduces to Eq. (23) of \cite{nz} for $B=0$. 

To find the lowest order of gravitational Faraday effect, we calculate the above integral (Eq. \ref{ki}) considering the weak-field and slow-rotation approximation, i.e., neglecting $(M/r)^1$ and $(a/M)^2$ \cite{nz}, and rewrite it as
\begin{eqnarray}\nonumber
 \chi &=& -\int_{r_{\rm orb}(\th)}^{r_o} \int_{\th_s}^{\th_o} \left[\f{2aM^2}{r^4}-\f{3aMB^2}{2r}(1+3\cos2\th) \right] \sin\th dr d\th 
 \\
&=& \int_{r_{\rm orb}(\psi)}^{r_o} \int_{-\psi_0}^{\psi_0} \left[\f{2aM^2}{r^4}-\f{3aMB^2}{r}(3\psi^2-1) \right] dr d\psi
 \label{ki2}
\end{eqnarray}
where we substitute $\psi=\cos\th$ in the last expression, and hence, the integration limit is changed. See Appendix \ref{app2} for the discussion on the integration limit. Performing the integration over $r$, we obtain from Eq. (\ref{ki2})
\begin{eqnarray}
 \chi &=& \int_{-\psi_0}^{\psi_0} \left[\f{2aM^2}{3}\left(\f{1}{(r_{\rm orb}(\psi))^3}-\f{1}{r_o^3}\right)-3aMB^2(3\psi^2-1){\rm ln}\left(\f{r_o}{r_{\rm orb}(\psi)}\right) \right] d\psi
 \label{ki3}
\end{eqnarray}
where (see Eq. \ref{b7}), 
\begin{eqnarray}
 r_{\rm orb}(\psi)=\f{r_{\rm min}}{\sqrt{1-(r_{\rm min}^2/\eta)\psi^2 }}.
\end{eqnarray}
Finally, we obtain from Eq. (\ref{ki3})
\begin{eqnarray}\nonumber
 \chi = aM^2\cos\th_0 \left(\f{\pi}{4r_{\rm min}^3}-\f{4}{3r_o^3}\right)-6aMB^2\cos\th_0\left[1-\f{4}{3}\cos^2\th_0 - \sin^2\th_0~ {\rm ln}\left(\f{2r_o}{r_{\rm min}}\right)   \right],
 \\
 \label{ki4}
\end{eqnarray}
substituting $\psi_0=\sqrt{\eta}/r_{\rm min}=\cos\th_0$.
Eq. (\ref{ki4}) reduces to the expression obtained earlier \cite{nz} for $B \rightarrow 0$ and $r_o \rightarrow \infty$. We do not consider $r_o \rightarrow \infty$ unlike \cite{nz}, as the value of $r_o$ gives a finite correction for the logarithm term obtained in Eq. (\ref{ki4}) due to the presence of a non-zero $B$. One can notice that Eq. (\ref{ki4}) vanishes for the orbits in the equatorial plane ($\th_0 = \pi/2$), which commensurate to Sec. A of \cite{nz}. Therefore, no gravitational Faraday effect can be seen for the equatorial orbits, even in the presence of a non-zero magnetic field. In addition, the effect of magnetic field on the gravitational Faraday rotation is absent for 
\begin{eqnarray}
 \th_0=\sin^{-1}\left(\f{1}{\sqrt{4-3{\rm ln}\left(\f{2r_o}{r_{\rm min}}\right)}} \right),
\end{eqnarray}
as the terms inside the square bracket of Eq. (\ref{ki4}) vanish. 
In case of the slowly rotating black hole $(a/M << 1)$ and for a distant observer ($M/r << 1$), one can obtain (using Eq. \ref{fra} or Eq. \ref{ki4}) the separation angle ($\T$) of the right and left circularly polarized beams arisen due to the gravitational Faraday rotation  as
\begin{eqnarray}\nonumber
 \T|_{(a/M << 1, M/r << 1)}=\pm \f{a}{\o}\left\{M\cos\th_0 \left(\f{\pi}{4r_{\rm min}^3}-\f{4}{3r_o^3}\right)-6B^2\cos\th_0\left[1-\f{4}{3}\cos^2\th_0 - \sin^2\th_0~ {\rm ln}\left(\f{2r_o}{r_{\rm min}}\right)   \right]\right\} .
 \\
 \label{ki6}
\end{eqnarray}
Eq. (\ref{ki6}) reveals, although the gravitational Stern-Gerlach effect arises due to the inhomogeneous gravitomagnetic field (as it depends on the position), it is also affected by the constant magnetic field $B$. The first term ($\propto r^{-3}$) is already reported in several papers \cite{fl, nz, ish, ser, mas1, mas2, fr1, fr2} but the second term, i.e., the logarithmic correction in the gravitational Faraday rotation and Stern-Gerlach effect due to the presence of magnetic field is completely new.

Note that Eq. (\ref{cgk}) reduces to the combinations of the expressions of $B_g^r$ and $B_g^{\th}$ of Sec. V.B of Ref. \cite{nz} for $B \rightarrow 0$. There is no discrepancy between Eq. (\ref{cck}) of this paper and Eq. (131) of \cite{fr1} for $B \rightarrow 0$. If one multiplies Eq. (\ref{cck}) with $h^{3/2}$, it reduces to Eq. (131) of \cite{fr1}. The apparent discrepancy between these two expressions arises due to the different definition of the $\g_{ij}$ metric (see Eq. \ref{gmij} of this paper and Eq. 6 of \cite{fr1}). Moreover, Eq. (131) of \cite{fr1} derived $\left({\rm curl}~{\rm curl}~{\bf g}\right)$ whereas we derive $\left({\rm curl}~(\sqrt{h}~{\rm curl}~{\bf g})\right)$. Eventually, the term inside the square bracket of Eq. (\ref{ki}) reduces to Eq. (23) of \cite{nz} for $B \rightarrow 0$, as the final expression (scalar quantity) for calculating the separation angle ($\T$) due to the gravitational Stern-Gerlach effect.

\section{\label{sec4}Gravitational Faraday rotation in the magnetized Reissner-Nordstr\"om spacetime}

\subsection{\label{sec4.1}Brief discussion on the Reissner-Nordstr\"om spacetime immersed in the uniform magnetic field}

If the Reissner-Nordstr\"om (RN) spacetime is immersed in a uniform magnetic field $B$, the transformed metric can be expressed as \cite{er}
\begin{eqnarray}
 ds^2=|\L|^2 \left[\f{\D}{r^2}~dt^2-\f{dr^2}{\f{\D}{r^2}}-r^2d\th^2 \right]-|\L|^{-2} r^2\sin^2\th\left( |\L_0|^2d\phi-\varpi dt \right)^2
 \label{rn}
\end{eqnarray}
where
\begin{eqnarray}
\D=r^2-2Mr+Q^2\,\, , \,\, \,\, \,\,
 \L=1+\f{1}{4}B^2\left(r^2\sin^2\th+Q^2\cos^2\th \right)-iBQ \cos\th
\end{eqnarray}
and 
\begin{eqnarray}
 \varpi=\f{BQ}{2r}\left[-4 + 2B^2r^2 + B^2Q^2 - 
 B^2 \D \sin^2\th \right].
 \label{o}
\end{eqnarray}
Here, $M$ and $Q$ are the mass and charge of the spacetime which is immersed in a uniform magnetic field $B$. The mRN spacetime will reduce to the ordinary RN spacetime (oRN) for $B=0$. One should note here that the mRN metric (Eq. \ref{rn}) is the exact electrovacuum solution of the Einstein-Maxwell equation (with a non-zero magnetic field), similar to the oRN. Another similarity between oRN and mRN is that the locations of the horizons ($r_{\pm}$) occur at the same distance, i.e., 
\begin{eqnarray}
 r_{\pm}=M\pm \sqrt{M^2-Q^2}
 \label{rh}
\end{eqnarray}
where $r_+$ and $r_-$ indicate the locations of the event horizon and Cauchy horizon, respectively. Eq. (\ref{rh}) shows that both of the horizons exist only for $ 0 < Q \leq M$, whereas the horizons vanish for the overcharged ($Q > M$) RN spacetime. The reason to add 
\begin{eqnarray}
  |\L_0|^2=|\L(r,0)|^2=\left(1+\f{B^2Q^2}{4} \right)^2+B^2Q^2
\end{eqnarray}
before $d\phi$ of Eq. (\ref{rn}) is same as discussed in Sec. \ref{sec3.1}. Thus, we do not repeat it here. Now, the Cartan components of the electric $(E)$ and
magnetic fields ($H$) in the mRN spacetime are given by (see Eq. 4.5 and Eq. 4.6 of \cite{er}):
\begin{eqnarray}
 H_r+iE_r&=&\L^{-2}\{iQ/r^2\left[1-\f{1}{4}B^2\left(r^2\sin^2\th+Q^2\cos^2\th \right)\right] \nonumber
 \\
 &+& B(1-1/2iBQ\cos\th)(1-Q^2/r^2)\cos\th \} ,
 \\
 H_{\th}+iE_{\th}&=&-B\L^{-2}(1-1/2iBQ\cos\th)\f{\sqrt{\D}}{r}\sin\th.
\end{eqnarray}

Thus, ${\bf E} \times {\bf H}$ serves as the source of twist potential in the mRN spacetime as

\begin{eqnarray}
 |{\bf E} \times {\bf H}|=|E_r H_{\th}-E_{\th} H_r|=-\f{BQ\sqrt{\D}\sin\th}{16r^3 \L^4}~\left[1-\f{1}{4}B^2\left(r^2\sin^2\th+Q^2\cos^2\th \right)\right].
\end{eqnarray}
If $B$ and/or $Q$ vanishes, the twist potential becomes zero. 
The stark contrast between an oRN and a mRN is that the mRN is an stationary and axisymmetric spacetime whereas the oRN is a static and spherically symmetric spacetime. Which makes this important difference between these two? Of course, the presence of magnetic field. If the magnetic field vanishes, $\varpi$ becomes zero. One can see the non-zero gravitational Faraday rotation or the so-called spin precession only because of the presence of $\varpi$ or $g_i$ or $B$. 
It is interesting to see that $\varpi$ or $g_i$ can vanish at 
\begin{eqnarray}
r_0=r|_{\varpi=0}=\f{\sqrt{8-2 B^2 Q^2+\left(3 B^2 Q^2-4\right) \sin^2\th+B^2 (M^2-Q^2)
\sin^4\th}-MB \sin^2\th}{B \left(2-\sin^2\th\right)}
\label{r0}
\end{eqnarray}
for a particular radius $r=r_0$ even if $B \neq 0$. So, one cannot see the gravitational Faraday rotation at this particular orbit whereas the ordinary Faraday rotation can be seen. Note that to obtain a physically realistic $r_0$, it must be positive, which satisfies 
\begin{eqnarray}
 \sin^2\th \geq \left(1-\f{4}{B^2Q^2}\right)\, ,\,\,\,\, {\rm i.e,} \,\,\,\,\, BQ \leq 2\sec\th.
\end{eqnarray}

\subsection{\label{sec4.2}Gravitational Faraday rotation and Stern-Gerlach effect in the magnetized Reissner-Nordstr\"om spacetime}
For the mRN spacetime, one can deduce 
\begin{eqnarray}
 g_{00}=h=\left(\f{\D |\L|^2}{r^2}-\f{r^2\varpi^2\sin^2\th}{|\L|^2}\right)\, , \,\,\,\,\, {\bf g} \equiv g_{\phi}=-\f{\varpi r^4 |\L_0|^2 \sin^2\theta}{\D|\L|^4-r^4\varpi^2\sin^2\th}
\end{eqnarray}
and
\begin{eqnarray}
\g_{ij}dx^idx^j=\f{r^2 |\L|^2}{\Delta}dr^2  
+r^2 |\L|^2 d\theta^2+\f{ \D |\L_0|^4}{h} \sin^2\th d\phi^2
\label{mgrn}
\end{eqnarray}
from Eq. (\ref{rn}). Now, using Eq. (\ref{mgrn}), we obtain
\begin{eqnarray}
 ({\rm curl}~{\bf g})^p &=& \left[\f{4Q \cos\th}{r^{1/2}(r-2M)^{1/2}}~ \d^p_{r} -\f{2Q \sin\th (r-4M)}{r^{3/2}(r-2M)^{3/2}}~ \d^p_{\th}+\mathcal{O}(Q^3) \right]B+\mathcal{O}(B^3)
\label{cgrnm}
\end{eqnarray}
and 
\begin{eqnarray}
\left({\rm curl}~(\sqrt{h}~{\rm curl}~{\bf g})\right)^p=\left[\f{4Q (r-3M)}{r^{5/2}(r-2M)^{3/2}}~\d^{p}_{\phi}+\mathcal{O}(Q^3)\right]B+\mathcal{O}(B^3).
\end{eqnarray}

Now, one can calculate the effective force for the gravitational Faraday rotation that varies as 
\begin{eqnarray}
 \left|h~{\rm curl} (\sqrt{h}~{\rm curl}~{\bf g}) \right| &=&\left[\f{4Q (r-3M)\sin\th}{r^{5/2}(r-2M)^{1/2}}+\mathcal{O}(Q^3)\right]B+\mathcal{O}(B^3)
  \label{fern}
  \\
  & \approx & \f{4BQ \sin\th}{r^2}
  \label{fern2}
 \end{eqnarray}
 where Eq. (\ref{fern2}) is valid far away $(M/r << 1)$ from the mRN spacetime. This indicates that the effective force is proportional to $r^{-2}$, as the magnetic field is non-zero here. This is similar to the `magnetic' correction term of the magnetized Kerr case.
 
  Now, one has to integrate the following expression 
 \begin{eqnarray}
 \chi &=& - 2BQ\int_{r_s}^{r_o} \int_{\th_s}^{\th_o} \f{(r-3M)\sin\th}{(r-2M)^2} dr d\th 
 \\
 &\approx & -2BQ \int_{r_s}^{r_o} \int_{\th_s}^{\th_o} \f{\sin\th}{r} dr d\th 
 \label{rnf0}
 \end{eqnarray}
 to find the lowest order (i.e., neglecting $(M/r)^1$ and $(Q/r)^2$) of gravitational Faraday effect for the magnetized RN spacetime. 
 Here we follow exactly the same procedure what we followed in Sec. {\ref{sec3.3}} to calculate Eq. (\ref{ki}), and obtain from Eq. (\ref{rnf0}),
  \begin{eqnarray}\nonumber
\chi &=& 2BQ \int_{-\psi_0}^{\psi_0} {\rm ln}\left(\f{r_o}{r_{\rm orb}(\psi)}\right)  d\psi
 \\
 &=& 4BQ\cos\th_0 \left[1-{\rm ln}\left(\f{2r_o}{r_{\rm min}} \right) \right].
 \label{rnfx}
\end{eqnarray}
Eq. (\ref{rnfx}) reveals that one cannot see the gravitational Faraday effect for the equatorial orbits in the mRN spacetime. However, the angular separation of the right and left circularly polarized beams due to the gravitational Faraday rotation is obtained using Eq. (\ref{fra}) or Eq. (\ref{rnfx}) as
\begin{eqnarray}
 \T=\pm \f{4BQ}{M\o}\cos\th_0 \left[1-{\rm ln}\left(\f{2r_o}{r_{\rm min}} \right) \right].
 \label{rnf}
\end{eqnarray}
The separation angle $(\T)$ depends on the logarithm of the observer's distance in this case too. This is similar to the magnetized Kerr case. It seems that the presence of a uniform magnetic field in a spacetime is responsible for $\T$ to depend on the logarithm of the distance of the observer. Interestingly, Eq. (\ref{rnf}) shows that $\T$ is inversely proportional to $M$ unlike the magnetized Kerr case. 

Note that an unmangnetized Reissner-Nordstr\"om spacetime cannot show the gravitational Faraday rotation and Stern-Gerlach effect, which leads to $\T=0$. On the other hand, a mRN spacetime shows the gravitational Faraday rotation, the gravitational Stern-Gerlach effect, the spin precession of a test gyro and the frame-dragging effect. In conclusion, the mRN spacetime acts as a perfectly rotating charged collapsed object, although its spin/Kerr parameter is zero. The source of its angular momentum is the Poynting vector or the non-zero twist potential $({\bf E}\times {\bf H})$. Recently, Ref. \cite{deri} has discussed the possibility of Faraday rotation even in Schwarzschild spacetime, due to the possibility of curvature-dependent interactions.

\section{\label{mls}Gravitational Faraday rotation and Stern-Gerlach effect in the massless charged Reissner-Nordstr\"om-like spacetime immersed in a uniform magnetic field }
If the mass term vanishes ($M \rightarrow 0$) in Eq. (\ref{rn}), it implies a massless charged RN-like spacetime \cite{vaa}. Now, if the massless charged RN-like spacetime is immersed in a uniform magnetic field $B$, the transformed metric can be expressed as (setting $M \rightarrow 0$ in Eq. \ref{rn}) 
\begin{eqnarray}
 ds^2=|\L|^2 \left[-\left(1+\f{Q^2}{r^2}\right)dt^2+\f{dr^2}{1+\f{Q^2}{r^2}}+r^2d\th^2 \right]+|\L|^{-2} r^2\sin^2\th\left( |\L_0|^2 d\phi-\varpi dt \right)^2
 \label{0rn}
\end{eqnarray}
where \cite{ag1, ag2}
\begin{eqnarray}
 \L&=&1+\f{1}{4}B^2\left(r^2\sin^2\th+Q^2\cos^2\th \right)-iBQ\cos\th ,
 \\
 |\L_0|^2&=&|\L(r,0)|^2=\left(1+\f{1}{4}B^2 Q^2 \right)^2+B^2Q^2
\end{eqnarray}
and
\begin{eqnarray}
 \varpi=\f{BQ}{2r}\left[-4 + 2B^2r^2 + B^2Q^2 - 
 B^2r^2\sin^2\th \left(1+\f{Q^2}{r^2}\right)\right].
\end{eqnarray}
As the horizons do not exist for the metric presented in Eq. (\ref{0rn}), it represents a naked singularity, in principle. 
Now, one can obtain
\begin{eqnarray}
 ({\rm curl}~{\bf g})^p &=& \left[\f{4Q \cos\th}{r}~ \d^p_{r} -\f{2Q \sin\th}{r^2}~ \d^p_{\th}+\mathcal{O}(Q^3)\right]B+\mathcal{O}(B^3)
\label{cgrn}
\end{eqnarray}
and 
\begin{eqnarray}
\left({\rm curl}~(\sqrt{h}~{\rm curl}~{\bf g})\right)^p=\left[\f{4Q}{r^3}~\d^{p}_{\phi}+\mathcal{O}(Q^3)\right]B+\mathcal{O}(B^3).
\end{eqnarray}
The effective force for the gravitational Faraday rotation varies as 
\begin{eqnarray}
\left|h~{\rm curl} (\sqrt{h}~{\rm curl}~{\bf g}) \right| =\left[\f{4Q \sin\th}{r^2}+\mathcal{O}(Q^3)\right]B+\mathcal{O}(B^3).
 \end{eqnarray}
Thus, the separation angle of the right and left circularly polarized beams arisen due to the gravitational Faraday rotation is obtained using Eq. (\ref{fra}) as
\begin{eqnarray}
 \T=- \varepsilon\int_{\rm source}^{\rm observer}2BQ\left(\f{\sin\th}{r} \right)dr d\th
 \label{tml}
\end{eqnarray}
Although we have, so far, used $\varepsilon=\pm (M\o)^{-1}$ (see Sec. \ref{faro}), the same cannot be used in Eq. (\ref{tml}), as $M=0$ in this particular case. However, by solving the wave equation with the magnetic potential (see Eq. 118 of Ch. 8 of \cite{ch}), one obtains a modified plane wave equation due to the presence of magnetic field \cite{cy}. From that, one may conclude: $\varepsilon=\pm (\o^2/B^2)^{-1}$, where $|\varepsilon| << 1$. As the magnetic energy ($\sim B^2$) is much less than the gravitational energy, the contribution of $(\o^2/B^2)^{-1}$ is negligible compare to $(M\o)^{-1}$, and, therefore it does not appear in case of the magnetized Kerr and mRN spacetimes. On the other hand, $M$ vanishes in the massless RN spacetime, and this particular term $\varepsilon=\pm (\o^2/B^2)^{-1}$ becomes important for this case. Note, without solving the wave equation, one may also deduce the term $(\o^2/B^2)^{-1}$ from the dimension analysis. However, eventually, we obtain from Eq. (\ref{tml}):
\begin{eqnarray}
 \T &=& \mp \f{2B^3Q}{\o^2} \int_{r_s}^{r_o} \int_{\th_s}^{\th_o} \left(\f{\sin\th}{r} \right)dr d\th
 \\
 &=& \pm \f{4B^3Q}{\o^2}~\cos\th_0 \left[1-{\rm ln}\left(\f{2r_o}{r_{\rm min}} \right) \right],
 \label{tml1}
\end{eqnarray}
which looks very similar to Eq. (\ref{rnf}). Eq. (\ref{tml1}) also depends on the logarithmic term, which is expected.

\subsection{\label{sec5.1}Spin precession in the massless magnetized Reissner-Nordstr\"om-like spacetime}
Using Eq. (\ref{s2}), one can obtain the exact expression for the spin precession in the massless RN spacetime as,
\begin{eqnarray}
  \bf{\O_s} & \approx & \f{BQ}{r}\left[-2\cos\th~\hat{r}+\sin\th~ \hat{\th} \right]+\mathcal{O}(B^3)Q+\mathcal{O}(Q^3)
  \label{lt1}
\\
  &= & r~|{\bf E}|~|{\bf B}|\left[2\cos(\pi-\th)~\hat{r}+\sin(\pi-\th)~ \hat{\th} \right] 
  \label{lt3}
 \end{eqnarray}
where $|{\bf E}| \sim Q/r^2$ is the modulus of the electric field. Although we have calculated the exact expression of the above expression (Eq. \ref{lt1}), here we write it upto the the order of $B^2$ and $Q^2$. To draw an analogy with the spin (LT) precession $({\bf \O}_{s}^{\rm Kerr})$ for a slowly rotating Kerr black hole \cite{jh, cm}
\begin{equation}
{\bf \O}_{s}^{\rm Kerr}=\f{1}{r^3}\left[3({\bf J}.\hat{\bf r})\hat{\bf r}-{\bf J}\right]=\f{J}{r^3}\left[2\cos\th~\hat{r}+\sin\th~\hat{\th} \right]
\label{kr}
\end{equation}
($J=|{\bf J}|$ is the angular momentum of the Kerr spacetime), we can rewrite Eq. (\ref{lt1}) or Eq. (\ref{lt3}) as
\begin{eqnarray}
{\bf \O}_s=j\left[2\cos(\pi-\th)~\hat{r}+\sin(\pi-\th)~ \hat{\th} \right] 
\label{lt4}
  \end{eqnarray}
 where $j=|{\bf j}|\sim r~|{\bf E}|~|{\bf B}|$ (i.e., ${\bf E}$ is orthogonal to ${\bf B}$ as well as ${\bf E} \times {\bf B}$ is orthogonal to ${\bf r}$
 \footnote{The angular momentum density (${\bf j}$) of an electromagnetic (EM) field is written as
\begin{eqnarray}
 \bf{j} \sim \bf{r} \times (\bf{E} \times \bf{B}). \nonumber
 \label{jem}
\end{eqnarray}}) is the modulus of the angular momentum density (${\bf j}$) of the electromagnetic field. It is clear from Eq. (\ref{lt4}) that the spin precession frequency for the EM field does not follow the inverse cube law of distance. Although $|{\bf \O_s}| \propto j$, the angle is differed by $(\pi-\th)$. This means that if the direction of angular momentum (${\bf j}$) of the electromagnetic field is separated by an angle $\th$ with the radial direction $\hat{r}$, the direction of spin precession is occurred separated by an angle $(\pi-\th)$ with $\hat{r}$. 

As the massless charged particles have not been observed yet \cite{lech, xs} in nature, Sec. \ref{mls} is devoted only for the theoretical purpose. Another problem is that the massless charged RN-like spacetime describes a naked singularity. Thus, as of now, this particular spacetime is completely unrealistic. However, one can notice the sole importance of the magnetic fields and electric charges for the gravitational Faraday rotation and Stern-Gerlach effect without the presence of mass and rotation parameter. Specifically, as all the derivations of Sec. \ref{sec5.1} is independent of mass, those might be important for the very small massive charged particles, like, electrons, protons etc. The possible applications of this special spin precession derived in this section as well as the gravitational analog of Faraday rotation and Stern-Gerlach effect may be studied in the laboratory applying it to the analog models of gravity \cite{vis, cgm, cm2}. For example, one may proceed with the same arrangement as mentioned in \cite{liao} with some modifications (i.e., submerging the whole system in a uniform magnetic field and adding charges of a `very low mass' at the center of the radial vortex) to visualize the above-mentioned effect.

\section{\label{con}Conclusion and discussion}

Our calculation has revealed a precise relation between the gravitational Faraday rotation, gravitational Stern-Gerlach effect and the spin precession of a test spin in a general stationary spacetime. We have applied this to derive the exact expressions of the above mentioned effects for the magnetized Kerr, magnetized Reissner-Nordstr\"om and magnetized massless Reissner-Nordstr\"om-like spacetimes, and shown that the logarithm correction of the distance of the source and observer in the gravitational Faraday rotation and Stern-Gerlach effect for the said spacetimes is an important consequence of the presence of magnetic field. Interestingly, we have shown that the spin precession frequency in the magnetized massless Reissner-Nordstr\"om-like spacetime, is proportional to the angular momentum density $(j)$ of an ordinary electromagnetic field, and it does not follow the inverse cube law of distance like the slowly-rotating Kerr black hole.

In the original Faraday effect at the flat spacetime, a plane polarized electromagnetic wave rotates by an angle $(\d\th_F)$ \cite{nz}:
\begin{eqnarray}
 \d\th_F=\f{2\pi e^3}{m_e^2 c^2 \o^2}\int_0^D n_e(s) B_{||}(s) ds
 \label{ofe}
\end{eqnarray}
where $n_e(s)$ is the density of electrons at each point $s$ along the path, $D$  is the length of the path where the light and magnetic field interact,
$B_{||}(s)$ is the component of the interstellar magnetic field in the direction of propagation at each point $s$ along the path, $e$ is the charge of an electron;
$c$ is the speed of light in the  vacuum;
$m_e$ is the mass of an electron and $\o$ is the frequency of light. In contrast, the gravitational Faraday effect for an ordinary slowly-rotating Kerr black hole, is proportional to $aM^2$ and inversely  proportional to $r^3$. In the presence of magnetic field, the modified term is proportional to $aMB^2$ and ${\rm ln}~r$. Note that the gravitational analog of Faraday effect occurs only when a light ray passes through the vacuum region outside a {\it rotating} strongly gravitating object like rotating black hole, rotating neutron star etc. It cannot be seen in our laboratory located in the {\it non-rotating flat spacetime} where we observe the ordinary Faraday effect. Comparing Eq. (\ref{ki4}) and Eq. (\ref{rnfx}) with Eq. (\ref{ofe}), one can find that the ordinary Faraday effect does not have any mathematical relation with the gravitational Faraday effect.

In case of the gravitational Stern-Gerlach effect for an ordinary slowly-rotating Kerr black hole, it is proportional to $aM$ and inversely  proportional to $\o r^3$ for a distant observer, whereas, in the presence of magnetic field, the modified term is proportional to $aB^2$ and ${\rm ln}~r$, and is inversely proportional to $\o$. We have shown that the gravitational Faraday rotation and the gravitational Stern-Gerlach effect become non-zero if the RN spacetime is immersed in a magnetic field. In a stark contrast with the magnetized Kerr spacetime, the gravitational Stern-Gerlach effect in the magnetized RN spacetime is proportional to $BQ$, ${\rm ln}~r$ and inversely proportional to $M\o$ for a distant observer. In case of the massless charged RN spacetime, it depends on $\sim B^3Q~{\rm ln}~r/\o^2$. 
Overall, we have shown that the magnetic field has a non-negligible effect on the gravitational Faraday rotation and the gravitational Stern-Gerlach effect. Although the massless charge(s) cannot be found in nature, our result could be applicable for a very light mass $(M \rightarrow 0)$ charged particle. Note, as the exact expressions of the gravitational Faraday rotation and Stern-Gerlach effect derived from our result are applicable to the strong gravity regime, it could be helpful to study the effects of magnetic field on the propagation of polarized photons of the rapidly rotating collapsed object. Another important point which emerges, is that the general spin precession formulation (see Sec. 1.10.1 of \cite{ns} and \cite{ckp}) is applicable to any spin (with a little modification, as mentioned in Sec. \ref{sec2.1}), i.e., the polarization vector of a particle whether it is massive (gyroscope) or massless (photon, graviton etc.).

The EHT collaboration has recently studied the polarization of the ring \cite{eht7} and the magnetic field structure near M87* \cite{eht8}. They have found that a part of the ring is significantly polarized \cite{eht7}, and the polarization is attributed to the Faraday rotation \cite{eht8}. They also estimated that the magnetic field strength is $B \sim 1-30$ G. As a Kerr black hole \cite{eht5}, it is not very unlikely that this resulting polarization of M87* could be a mixture of both of the usual Faraday rotation and the gravitational Faraday rotation with the additional contribution from the non-zero magnetic field, which we presented in Eq. (\ref{ki4}). Therefore, one may try to extract those contributions from the data and image released by the EHT collaboration for M87* in 2021 \cite{eht7, eht8}. Note, although the estimated magnetic field strength is not so high for M87*, a magnetic field of several hundred Gauss could be present near Sgr A* \cite{eatough}. Therefore, the gravitational Faraday effect due to the contribution from the magnetic field could also be higher in Sgr A* compared to M87*. This was also one of our motivations to study the gravitational Faraday effect in the magnetized Kerr spacetime.

In this paper, we have applied the modified geometric optics
developed in \cite{fr1}. The trajectories of the polarized photons are the null curves, which coincide with the null geodesic for $\o \rightarrow 0$. A deviation of the null rays from the null geodesic is controlled by the small parameter $\varepsilon$. The formalism presented in this paper has been developed from the point of view of the static
observer, and it is well applicable for a light ray passes through the vacuum region outside the rotating matter \cite{nz}, i.e., a null geodesic passes close to the black holes. Therefore, we have applied this formalism for the equatorial orbits and the symmetric orbits about it (following \cite{nz}), to find the lowest order of gravitational Faraday effect (arisen due to the presence of magnetic field) for those orbits. From the astrophysical point of view, our result provided in this paper is well applicable for the null geodesics which occur outside the ergoregion, i.e., our result covers a large space from ergoregion to infinity, in principle. For example, one can calculate the rotation $(\chi)$ of the plane of polarization of a light ray (from Eq. {\ref{ki4}}), which passes close to a Kerr black hole immersed in a magnetic field.

Our formalism is not applicable inside the ergoregion, as it diverges there. This is at least clear from Eq. (\ref{ki}) for the Kerr spacetime. Therefore, we need to modify the current formalism of the gravitational Faraday rotation and the gravitational Stern-Gerlach effect following the formalism developed in \cite{ckp}, which we plan to report soon in a different article. One has to modify this present formalism by attaching the photons to stationary observers that move with a nonzero angular velocity \cite{ckp}, which helps to avoid the divergence at the ergosurface. This approach would allow us to study the polarization-dependent effects for photons which closely approach a (non-)magnetized collapsed object.

It could be interesting to study how the polarization of light due to the gravitational Faraday rotation and/or the gravitational Stern-Gerlach effect modifies the black hole shadow of M87* \cite{eht7, eht8} in the presence of magnetic field. If the `bright' radiation behind the rotating collapsed object (such as, M87*) is non-monochromatic, the position of the shadow should
depend on the frequency ($\o$) of the radiation, i.e, one may observe a peculiar `rainbow effect' \cite{fr1} for the shadow of a collapsed object. The polarization splitting might also be detectable in future by the astrophysical observations. Finally, it would be worth to study the possible applications of the gravitational Faraday rotation and/or the gravitational Stern-Gerlach effect in the realistic astrophysical problems like \cite{bcb2, cbcr}.
\\

\section*{Acknowledgements} We thank the referee for constructive comments that helped to improve the
manuscript.

\appendix

\section{Projection of the orbit in the $(r, \th)$ plane for the
slowly-rotating Kerr spacetime\label{app2}}
The projection of the orbit in the $(r,\th)$ plane for Kerr metric is governed by the following equation (see Eqs. 178, 190, 191 of Chapter 7 of \cite{ch})
\begin{eqnarray}
 \int_r \f{dr}{\sqrt{r^4+(a^2-\xi^2-\eta)r^2+2M[\eta+(\xi-a)^2]r-a^2\eta}}=\int_{\th} \f{d\th}{\sqrt{\eta+a^2\cos^2 \th-\xi^2 \cot^2 \th}}
 \label{b1}
\end{eqnarray}
where $\eta$ and $\xi$ are constants of motion with $\eta > 0$ which corresponds to the null geodesics which intersect the equatorial plane and are symmetric about it \cite{ch, nz}. This equation determines the family of null geodesics
reaching the observer from an emitting ring. The above integration is performed for the weak deflections only, i.e., to find the lowest order of gravitational Faraday effect, we calculate the above integration for $a/r << 1$, $M/r << 1$ (following \cite{nz}) with the weakly magnetized collapsed objects. Therefore, Eq. (\ref{b1}) reduces to 
\begin{eqnarray}
 \int_r \f{dr}{\sqrt{r^4-(\xi^2+\eta)r^2}}=\int_{\th} \f{d\th}{\sqrt{\eta-\xi^2 \cot^2 \th}}.
 \label{b2}
\end{eqnarray}
Note, Eq. (\ref{b2}) does not include any term related to any of the so-called black hole hairs (see Appendix of \cite{nz}), i.e., $a$ and/or $M$, as long as the weak deflection (i.e., the lowest order of gravitational Faraday effect) is concerned. In a similar manner, Eq. (\ref{b2}) does not depend on $B$ and/or $Q$, and, hence the said equation is also applicable to the magnetized Kerr as well as magnetized Reissner-Nordstr\"om black holes which we consider in this paper. Eq. (\ref{b2}) mainly depends on two parameters, $\eta$ and $\xi$ which are in fact related to the `celestial coordinates' $\a$ and $\beta$ \cite{ch} of the image as seen by a distant observer who receives the light ray. One can readily verify that \cite{ch}
\begin{eqnarray}
 \a &=& \xi~{\rm cosec}~\th_0
\\
 \b &=& (\eta -\xi^2 \cot^2\th_0)^{1/2}
 \label{ab}
\end{eqnarray}
or, conversely,
\begin{eqnarray}
 \xi &=& \a \sin \th_0
\\
 \eta &=& \b^2+\a^2  \cos^2\th_0 ,
\end{eqnarray}
where $\th_0$ is the angular coordinate of the distant observer, $\a$ is the apparent perpendicular distance of the image from the axis of symmetry and $\b$ is the apparent perpendicular distance of the image from its projection on the equatorial plane \cite{ch}.

Now, performing the integration of the left hand side (LHS) in Eq. (\ref{b2}), one obtains 
\begin{eqnarray}
 \int \f{dr}{r^2\sqrt{1-r_{\rm min}^2/r^2}}=\f{1}{r_{\rm min}} \cos^{-1} \left(\f{r_{\rm min}}{r} \right)
 \label{b5}
\end{eqnarray}
where $r_{\rm min}=\sqrt{\xi^2+\eta} \equiv \sqrt{\a^2+\b^2}$ is the leading term of the largest root of the denominator of LHS of Eq. (\ref{b1}) for the small deflection. Substituting $\psi=\cos\th$, one can obtain from the right hand side (RHS) of Eq. (\ref{b2})
\begin{eqnarray}
 -\int \f{d\psi}{\sqrt{\eta-\psi^2 r_{\rm min}^2}}
 =-\f{1}{r_{\rm min}} \sin^{-1} \left(\psi \sqrt{\f{r_{\rm min}^2}{\eta}} \right).
 \label{b6}
\end{eqnarray}
Now, equating Eq. (\ref{b5}) and Eq. (\ref{b6}) we obtain
\begin{eqnarray}
 r_{\rm orb}=\f{r_{\rm min}}{\sqrt{1-(r_{\rm min}^2/\eta)\cos^2 \th}},
 \label{b7}
\end{eqnarray}
which is the projection of the orbit in the $(r,\th)$ plane for small deflections. In fact, no deflection is seen in this case \cite{nz}, as Eq. (\ref{b7}) does not depend on $M$, $a$, $Q$ and $B$. It is noticed that one obtains $\cos\th=\pm \sqrt{\eta}/r_{\rm min}$ for a very large $r$, i.e., $r \rightarrow \infty$. Here, the plus and minus signs correspond to the position angles $\th_o$ (i.e., $\psi_0=\sqrt{\eta}/r_{\rm min}$) and $\th_s$ (i.e., $\psi_s=-\sqrt{\eta}/r_{\rm min}=-\psi_0$) of the observer and source respectively.

\end{document}